\documentclass[12pt]{article}
\usepackage{epsfig,graphicx}

\title{Microscopic study of  the string breaking  in QCD}
\author{ A.M.Badalian, V.D.Orlovsky and Yu.A.Simonov,\\ Institute of Theoretical and Experimental
Physics\\ 117118, Moscow, B.Cheremushkinskaya 25, Russia}
\date{}

\newcommand{\be}{\begin{equation}}
\newcommand{\ee}{\end{equation}}

\def\la{\mathrel{\mathpalette\fun <}}

\def\fun#1#2{\lower3.6pt\vbox{\baselineskip0pt\lineskip.9pt
\ialign{$\mathsurround=0pt#1\hfil ##\hfil$\crcr#2\crcr\sim\crcr}}}

\newcommand{{\SD}}{\rm SD}

\newcommand{\vex}{\mbox{\boldmath${\rm x}$}}
\newcommand{\vey}{\mbox{\boldmath${\rm y}$}}
\newcommand{\ver}{\mbox{\boldmath${\rm r}$}}

\newcommand{\vep}{\mbox{\boldmath${\rm p}$}}
\newcommand{\veq}{\mbox{\boldmath${\rm q}$}}

\newcommand{\vel}{\mbox{\boldmath${\rm l}$}}
\newcommand{\veR}{\mbox{\boldmath${\rm R}$}}

\newcommand{\veu}{\mbox{\boldmath${\rm u}$}}
\newcommand{\vev}{\mbox{\boldmath${\rm v}$}}

\newcommand{{\Mc}}{\mathcal{M}}
\newcommand{\llan}{\langle\langle}
\newcommand{\rran}{\rangle\rangle}
\newcommand{\lan}{\langle}
\newcommand{\ran}{\rangle}

\begin{document}

\maketitle
\begin{abstract}
Theory of strong decays defines in addition to decay widths, also the
channel coupling and the mass shifts of the levels above the decay
thresholds. In the standard decay models of the $^3P_0$ type the
decay vertex is taken to be a phenomenological constant $\gamma$
and such a choice leads to large mass shifts of all meson levels
due to real and virtual decays, the latter giving a divergent
contribution. Here we show that taking the microscopic details of
decay vertex into account, one obtains new string width coefficient, which strongly suppresses virtual decay contribution.
In addition for a realistic space structure of the decay vertex of
highly excited states, the decay matrix elements appear to be
strongly different from those, where the constant $\gamma$ is used.
From our analysis also follows that so-called flattening potential
can imitate the effects of intermediate decay channels.

\end{abstract}

\section{Introduction}
There is   large variety of single-channel models, proposed decades ago, which
describe spectra of hadrons with reasonable accuracy \cite{1}. The most popular
and widely used is the relativistic quark model of N.Isgur and coworkers for
mesons \cite{2} and baryons \cite{3}, where effective constants are used for
quark masses (constituent masses), as well as an overall negative constant
$(C<0)$, and several additional parameters for spin-dependent interactions. For
heavy quarkonia the Cornell model \cite{4}, based on nonrelativistic
Schr\"{o}edinger equation and linear plus Coulomb potential, was extensively
exploited.

Most of the models proposed are rather successful in predictions of low-lying
hadron masses and the idea, that  relativistic quark Hamiltonian with confining
and the gluon-exchange potential, can be derived from QCD,  seems to be
realistic. It was indeed done in Ref.~\cite{5}, using the  Wilson loop and
field correlator technic, where for quarks at the ends of the rotating QCD
string the relativistic string Hamiltonian (RSH) was derived. The RSH contains
several improvements over old models:

i) At small $L$ (low rotation) it reduces to standard relativistic quark
Hamiltonian \cite{1,2,5,5'}, but with current quark masses used instead of
phenomenological constituent masses. The resulting hadron masses calculated are
exprssed through the former and the string tension $\sigma$ \cite{5,5',6}.

ii) The overall negative constant is absent while for a given
quark the universal negative self-energy correction appears,
calculated via $\sigma$ \cite{7}; its presence is crucially
important to reproduce linear behavior of the Regge trajectories.

iii) At high $L$ due to string rotation term, which naturally
appears in RSH \cite{5} and is absent  in quark Hamiltonian models
\cite{1,2,3,4}, the Regge trajectories with correct slope and
intercept are calculated \cite{8,9}

As a result, one  obtains  the formalism, derived from QCD with minimal number
of the first-principle parameters (current quark masses, $\alpha_s$, and string
tension $\sigma$; connection of two latter was found in \cite{10}).

Theoretical calculation of hadron masses with the use of RSH in single-channel
approximation was successful for all states below open decay thresholds (see
\cite{11} and \cite{12} for charmonium and bottomonium, \cite{13} for
heavy-light mesons, \cite{8,9,14} for light mesons, and \cite{15} for higher
pionic states). However, for states above threshold  RSH gives somewhat higher
masses  and one can expect that taking coupling to decay channels into account
one obtains mass shifts of these levels down, closer to experimental values.

To this end the channel-coupling (CC) models were formulated in
\cite{4}, \cite{16,17}. They are  based on the presumed form of
the decay Hamiltonian, which is usually taken to be the $~^3P_0$
model \cite{18,19}. More forms have been investigated in
\cite{20}, with a conclusion that the so-called $sKs$ model yields
results close to that within the  $~^3P_0$ one. Influence of the
CC effects on the spectrum are significant and can be divided in
two parts.

First, the effect of close-by channels, when the energy of the
level in question is not far from the two-body threshold (e.g.
$\psi(3770)$ in connection with $DD, DD^*$ thresholds). As was
found in \cite{16, 21}, the overall shift from the sum of the
nearest thresholds (e.g. for charmonium) is of the order of
(100-200) MeV.

Another part of the mass shift is associated with the contribution
from higher intermediate thresholds (e.g. of a pair of higher $D$
and $\bar D$ mesons) and in this case convergence of such terms
appears to be questionable. This topic was investigated in
\cite{22} and in the first paper of \cite{22} the authors have
introduced additional form factor for quarks to make the $~^3P_0$
vertex nonlocal and ensure convergence of the sum of contributions
over thresholds.

Therefore the structure of the string-breaking vertex becomes a fundamental
issue and one should try to find its properties from the basic QCD Lagrangian,
which takes into account both confinement and chiral symmetry breaking. Such
the strong decay Hamiltonian was derived from the first principles in \cite{24}
(which also supported  by the $sKs$  model in its   relativistic version), the
interaction kernel being simply the confining potential between the newly born
quark (antiquark $\bar q$) and original (possibly heavy)  antiquark $\bar Q$
(quark $Q$). This constitutes the strong decay term in action of the form  (in
the local limit  cf \cite{20})

\be S_{eff} = \int d^4 x \bar \psi_{\bar q} (x) \mathcal{M} (x)
\psi_q (x),\label{1}\ee

\be  \mathcal{M} (\vex, \vex_Q, \vex_{\bar Q}) =\sigma ( | \vex_q
- \vex_{\bar Q} | + | \vex_{\bar q} - \vex _Q |). \label{2}\ee

From (\ref{1}), (\ref{2}) one obtains the decay matrix element
between the original state $(Q\bar Q)_{n}$ and decay products --
two mesons $(Q \bar q)_{n_2}$ and $(\bar Q q)_{n_3}$ with relative
momentum $\vep$,

\be J_{nn_2n_3} (\vep) = \frac{1}{\sqrt{N_c}} \int \bar y_{123}
\Psi^{(n)}_{Q\bar Q} (\veu-\vev) e^{i\vep\ver} \mathcal{M}
(\vex, \veu, \vev) \Psi^{(n_2)}_{Q\bar q} (\veu-\vex)
\Psi^{(n_3)}_{\bar Q q} (\vex-\vev) d^3\vex d^3
(\veu-\vev).\label{3}\ee

Here the factor $\bar y_{123}$ accommodates spin-angular variables
and the functions in (\ref{3}) refer to radial dependencies only,
$\ver = c (\veu-\vev)$, $c\approx 1$ for heavy $Q\bar Q$ masses.
In the way  it was derived in \cite{24}, the $\mathcal{M} (\vex,
\veu, \vev)$ refers to the string between positions $\veu$ of
quark $Q$ and $\vev$ of antiquark $\bar Q$, which breaks at the
point $\vex$ somewhere between $\veu$ and $\vev$. It is clear,
that the point $\vex$ should lie in the body of the string, i.e.
within the string width $d$ from the axis of the string, see
Fig.1. This implies the necessity of an extra factor in (\ref{1}),
$\Theta_{\rm string}(\vex,\vel)$, which is proportional to the
energy density of the string with a fixed axis $\vel$ (the vector
$\veu-\vev$ in our case).

\begin{figure}[h]
\center{
\includegraphics{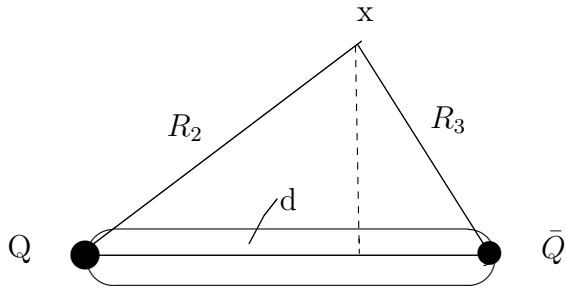}
\caption{ String breaking (pair creation) point $x$ and heavy-light radii $R_2$
and $R_3$, shown together with the string of radius $d$ between charges $Q$ and
$\bar Q$.}}
\end{figure}

Now the string density was studied both analytically \cite{25} and
on the lattice \cite{26,27}. In the  field correlator method
\cite{28,25} the string width $d$ is  proportional to small vacuum
correlation  length, $\lambda\approx 0.1 $ fm \cite{10,29}, and
therefore it is also small, $d\la 0.3 $ fm, for not highly
excited hadrons.

The string field density was computed in \cite{25,26,27} and one
can visualize there the field distribution in the string,
exponentially decreasing far from the string axis. In lattice
calculations similar estimates hold, but they depend on the way of
probing the string fields: in case of a connected probe one has
$d_{\rm con} \approx  0.3$ fm \cite{26} and in the case of a
disconnected probe $d$ is smaller, $d_{\rm disc}\approx 0.15$ fm
\cite{27}. A simple look into the configuration of large closed
Wilson loop for the string and a smaller one for $q\bar q$ closed
trajectory, shows that $d_{\rm disc}$ is closer to the string
breaking situation. In what follows we shall take $d$ to be
somewhere between the two (lattice) values. In next Section we
shall study the effect of the decaying string width, called the
factor $\Theta_{\rm string} (\vex,\vel)$, on the decay matrix
element and resulting mass shifts of energy levels.

\section{The width-of-the string correction in the string-breaking action}
In \cite{24} it was shown that the effective action of the $q\bar
q$ pair emission in the field of static charges $Q\bar Q$, placed
at fixed points, can be written as

\be S_{eff} = \int d^4 x d^4 y \bar \psi (x) \tilde  \mathcal{M}
(x,y) \psi(y),\label{4}\ee where the mass operator $ \tilde
\mathcal{M} (x,y)$ is to be found from the nonlinear (integral)
equation
\be
\tilde \mathcal{M} (x,y) = \left( \frac{1}{\hat
\partial + m_q + \tilde\mathcal{M}}\right)_{ (x,y)}  J(x,y)\label{5}\ee
and the kernel $J(x,y) = J_Q(x,y) + J_{\bar Q} (x,y)$ accounts for
the fields in the string. Taking into account only colorelectric
fields of scalar confining correlator $D(x)$, one can present
$J_Q(x,y)$ as

\be J_Q (x,y) = \frac{g^2}{N_c} \lan A_4 (x) A_4 (y)\ran = \int_Q^x du_i
\int^y_Q dv_i D(u-v).\label{6}\ee

Here $Q$ is the position  of the closest static charge $Q$( or $\bar Q$) in 4d
space and the analogous term appears  for the anticharge $\bar Q$ (or $Q$).
Note, that in \cite{24} it was tacitly implied that in (\ref{6}) the averaging
is over the vacuum configurations, and the points $\vex, \vey$ can be anywhere
in the space, surrounding static charge. It is the property of the kernel
$J(x,y)$ that it is asymptotically large for collinear $\vex||\vey$ , but the
direction of this vector can be arbitrary. That was enough for the proof of
Chiral Symmetry Breaking (CSB) due to confinement, but in our case one needs a
further specification.

Namely, at the moment of creation the created pair $q\bar q$ must lie on the
minimal surface of the Wilson loop of static charges $Q\bar Q$, i.e. on (or
inside) the string connecting static charges. This means  that we must replace
in (\ref{6}) $\lan A_4A_4\ran$ by $ \lan A_4(x) A_4(y)\ran_{\rm string}$, where
the latter acquires the string profile factor $\Theta_{\rm string} (x,y)$,
proportional to the string density of colorelectric fields, \be \Theta_{\rm
string} (x,y) = \xi (\vex, \vel) \xi (\vey, \vel).\label{7}\ee

Here $\vel = \vex_Q (t) -  \vex_{\bar Q} (t)$ is the string axis vector. For
long string, $|\vel|\gg \lambda$, one expects that $\xi$ depends only on the
distance $\vex_\bot$ from the string axis, e.g. \be \vex^2_\bot = \frac{|(\vex
- \vex_Q) \times \vel|^2}{\vel^2}.\label{8}\ee To simplify matter and for rough
estimates one can take $\xi$ as a Gaussian function of distance to the center
of the string, so we take \be \xi (\vex, \vel) \approx \exp \left( -\rho^2
\left(\vex- \frac{\vex_Q+\vex_{\bar Q}}{2}\right)^2\right),\label{9}\ee where
$\rho\sim \frac{1}{d} \sim O(1$ GeV).

Insertion of $\Theta_{\rm string} (x,y)$  in its local form, $\Theta_{\rm
string} (\vex,\vex)$, into (\ref{3}) is easily integrated and yields for
intermediate mesons with almost equal radius, $R_2\approx R_3$ (corresponding
SHO parameters $\beta_2 =\beta_3, \beta_i\sim 1/R_i$) \be J_{nn_2n_3} (\vep)
\to J_{nn_2n_3}^{\rm (string)}(\vep) = \eta (\beta_2, \rho)
J_{nn_2n_3}(\vep).\label{10}\ee Here \be \eta(\beta, \rho) =\left(
\frac{\beta^2_2}{2\rho^2 + \beta^2_2}\right)^{3/2}\cong \left(\frac{d^2}{d^2+
R^2_2}\right)^{3/2}.\label{11}\ee

One should note that the expression (\ref{11}) is valid as an
asymptotic estimate for large distances $R_2$, $R_2\gg d$. Besides
the approximation (\ref{9}) does not take into account an
additional suppression in the case of short strings, $|\vel|<
R_2+R_3$.

Thus $\eta(\beta, \rho)$ plays the role of the suppression factor for high
excited  intermediate mesons. Indeed, radii of high excited mesons are growing
with radial and orbital quantum numbers $(n,L), R_2(n,L) \sim \sqrt{L},
\sqrt{n}$, and $\eta^2(\beta, \rho) \sim \frac{1}{L^3}, \frac{1}{n^3}$.

\section{Study of the decay vertex}

The string decay vertex, derived in Ref.~\cite{24}, has the form
(\ref{2}) in the local approximation. In the standard $^3P_0$
approach \cite{19} it is assumed that one can effectively replace
the kernel $\mathcal{M} (\vex, \vex_Q,\vex_{\bar Q})$ by a
constant. The same type of approximation was used in \cite{28*,
29*, 30*} and also in \cite{24}, where results were compared with
the analysis of decays of $\psi(3730)\to D\bar D$ and $\Upsilon
(4S) \to B\bar B$ in \cite{31*}. Below we shall study the
reliability of this replacement and show that the replacement of the kernel
$\mathcal{M}$ by a constant is not always valid, especially for high excited states.

To illustrate this statement we consider the decay matrix element
(\ref{3}) with kernel $\mathcal{M}$, written in momentum space, where the wave functions of heavy-light mesons are replaced by gaussians $\Psi^{(n_2,n_3)}(q) = \left(\frac{2\sqrt{\pi}}{\beta_2}\right)^{3/2}e^{-q^2/2\beta_2^2}$ ($\beta_2=0.48$ for $D$-mesons and $\beta_2=0.49$ for $B$-mesons):
\begin{equation} \label{J(p,sigma)}
J_{nn_2n_3}(\textbf{p}) = \frac{\sigma}{\sqrt{N_c}} \frac{32\sqrt{2}\pi}{\beta_2^4} \int \bar{y}_{123}\Psi^{(n)}_{Q\bar{Q}}(c\textbf{p}+\textbf{q})e^{-q^2/\beta_2^2} \Phi\left(-\frac{1}{2}; \frac{3}{2}; \frac{q^2}{2\beta_2^2}\right) \frac{d^3q}{(2\pi)^3}.
\end{equation}
Here $\Phi(a;b;z)$ is the confluent hypergeometric function,
\begin{equation}\label{hypergeom}
\Phi(a;b;z) = 1+ \frac{a}{b}z+\frac{a(a+1)}{b(b+1)}\frac{z^2}{2!} + \ldots,
\end{equation}
the wave function $\Psi^{(n)}_{Q\bar{Q}}(c\textbf{p}+\textbf{q})$ can by expressed as a series of oscillator wave function (see Appendix 2 of \cite{30*} for details).

Another expression, which should be compared with (\ref{J(p,sigma)}), can be obtained by replacement of the kernel $\mathcal{M}$ to some constant $M_\omega$:
\begin{equation}\label{J(p,Momega)}
J^{(M_\omega)}_{nn_2n_3}(\textbf{p}) = \frac{M_\omega}{\sqrt{N_c}} \left(\frac{2\sqrt{\pi}}{\beta_2}\right)^3 \int \bar{y}_{123}\Psi^{(n)}_{Q\bar{Q}}(c\textbf{p}+\textbf{q})e^{-q^2/\beta_2^2} \frac{d^3q}{(2\pi)^3}.
\end{equation}

 We consider $nS$ ($n$ ranges from $1$ to $5$) charmonium states ($\Psi^{(n)}_{Q\bar{Q}}$ in (\ref{J(p,sigma)}) and (\ref{J(p,Momega)})), while the final states are $DD$ (or $DD^*, D^*D^*$) in all cases, then $\bar{y}_{123}$ is proportional to $q_i$. It is of interest
to notice that while for the $1S$, $2S$ and $3S$ states one can
reproduce $\mathcal{M}$ by constant rather well, on the
contrary, for the $4S$ and $5S$ states such the replacement
does not work (see left parts of Figs. 2,3,4,5,6). The constants $M_\omega$ appear to be different in these cases:
$M_\omega = 0.65$ GeV, 0.8 GeV and 1.1 GeV for the $1S, 2S$ and $3S$ states.

\begin{figure}[h]
\center{
\includegraphics[width= 6cm,height=5cm,keepaspectratio=true]{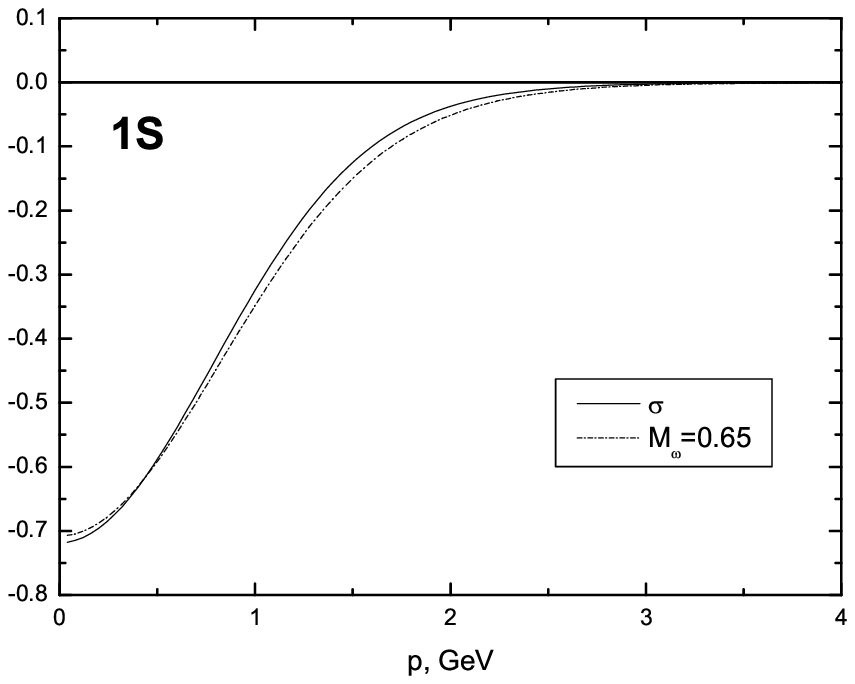}
\includegraphics[width= 6cm,height=5cm,keepaspectratio=true]{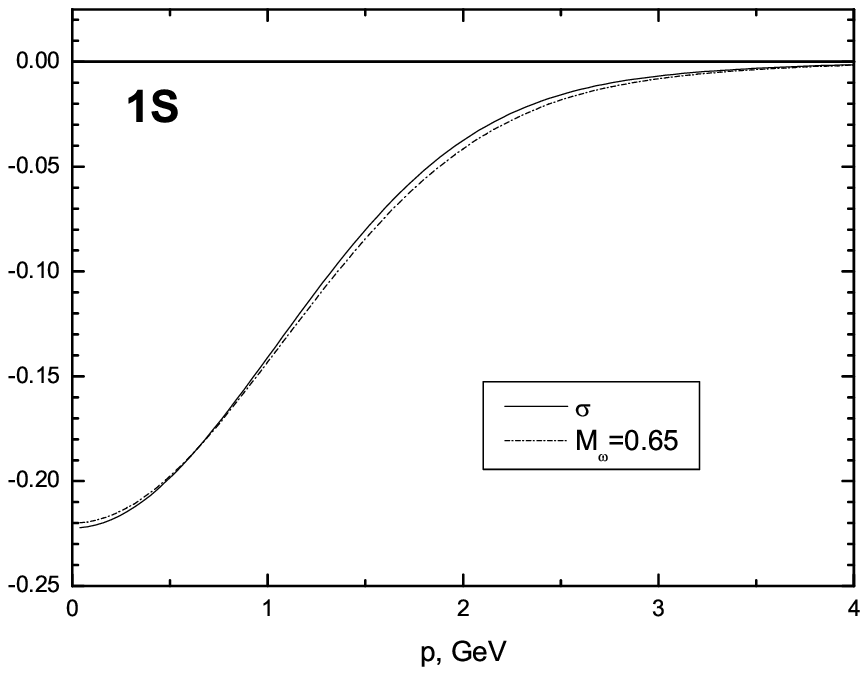}
\caption{Profiles of decay matrix elements, Eq.(3) (scalar parts), for $1S (c\bar c)$ into
$DD$ (left panel) and $1S(b\bar b)$ into $BB$ (right panel) calculated with
decay vertex of Eq. (2), expression (\ref{J(p,sigma)}) -- solid line, and for the constant decay vertex, expression (\ref{J(p,Momega)}) -- broken line.}}
\end{figure}

Surprisingly, that for the $nS$ bottomonium states the constant decay vertex reproduces results with the kernel $\mathcal{M}$ very well even for $4S$ and $5S$ states, however constants are different for different bottomonium states: $M_\omega$ varies from 0.65 GeV for $1S$ state to 1.3 GeV for $5S$ state (see right parts of Figs. 2,3,4,5,6), what we see in the case of charmonium states too.

\begin{figure}[h]
\center{
\includegraphics[width= 6cm,height=5cm,keepaspectratio=true]{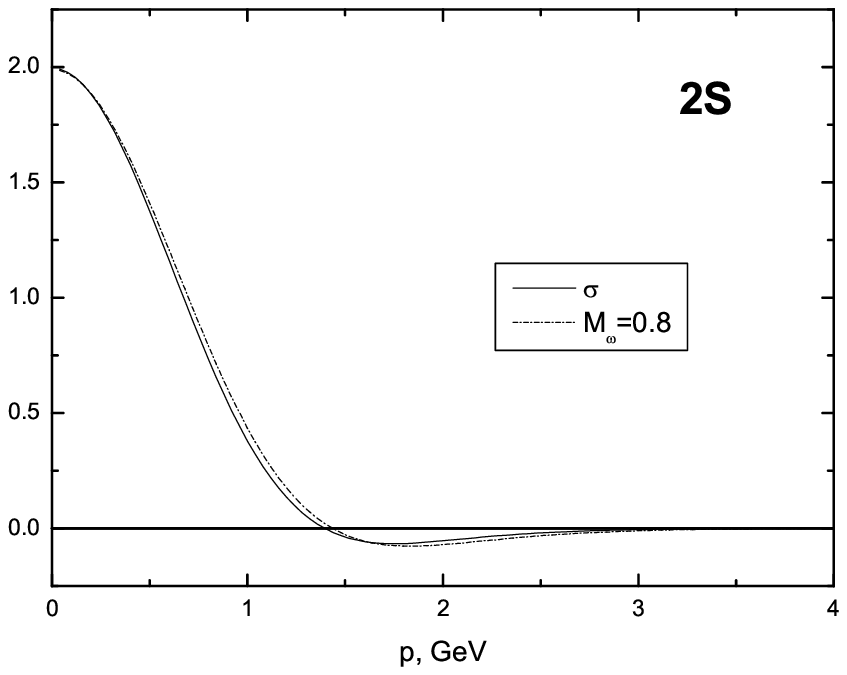}
\includegraphics[width= 6cm,height=5cm,keepaspectratio=true]{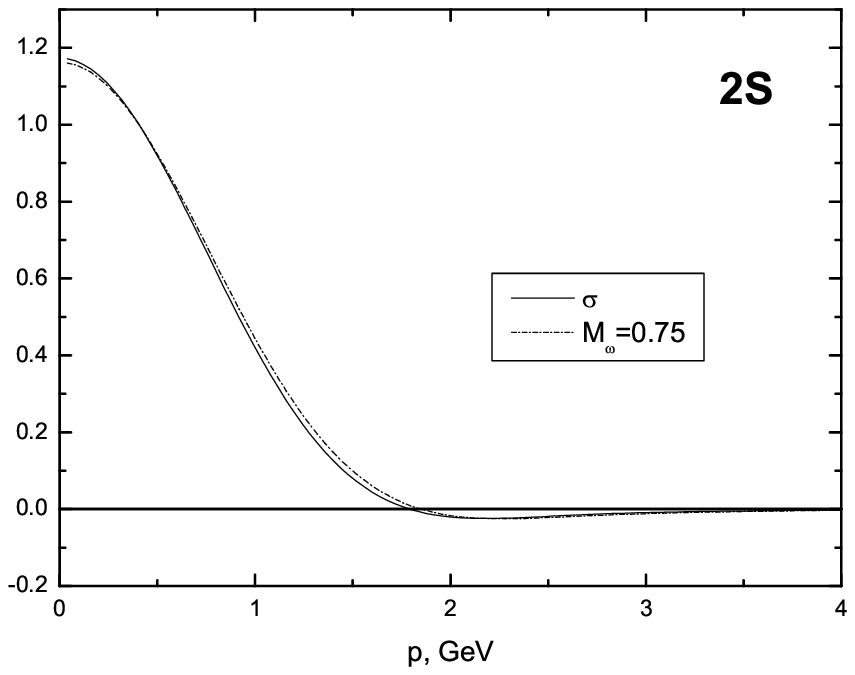}
\caption{The same as in Fig. 2, but for the  $2S (c\bar c)$ into $DD$ (left
panel) and $2S(b\bar b)$ into $BB$ (right panel).}}
\end{figure}

\begin{figure}[h]
\center{
\includegraphics[width= 6cm,height=5cm,keepaspectratio=true]{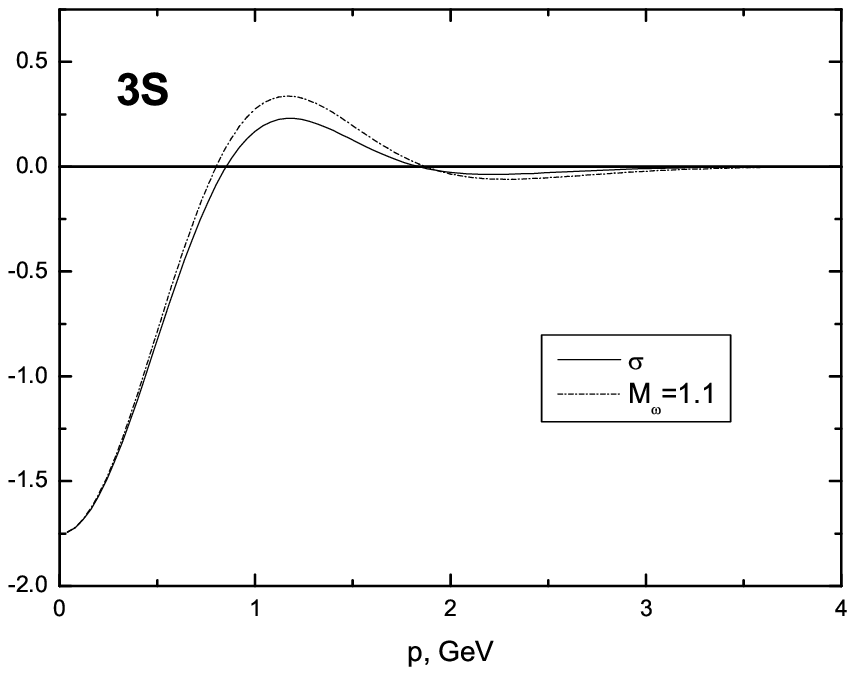}
\includegraphics[width= 6cm,height=5cm,keepaspectratio=true]{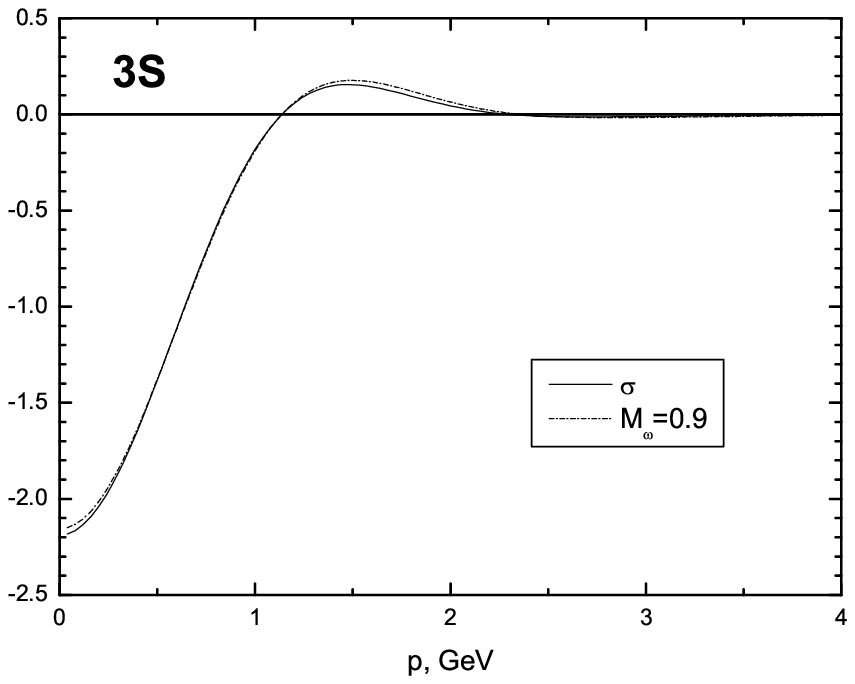}
\caption{The same as in Fig. 2, but for the  $3S (c\bar c)$ into $DD$ (left
panel) and $3S(b\bar b)$ into $BB$ (right panel).}}
\end{figure}

\begin{figure}[h]
\center{
\includegraphics[width= 6cm,height=5cm,keepaspectratio=true]{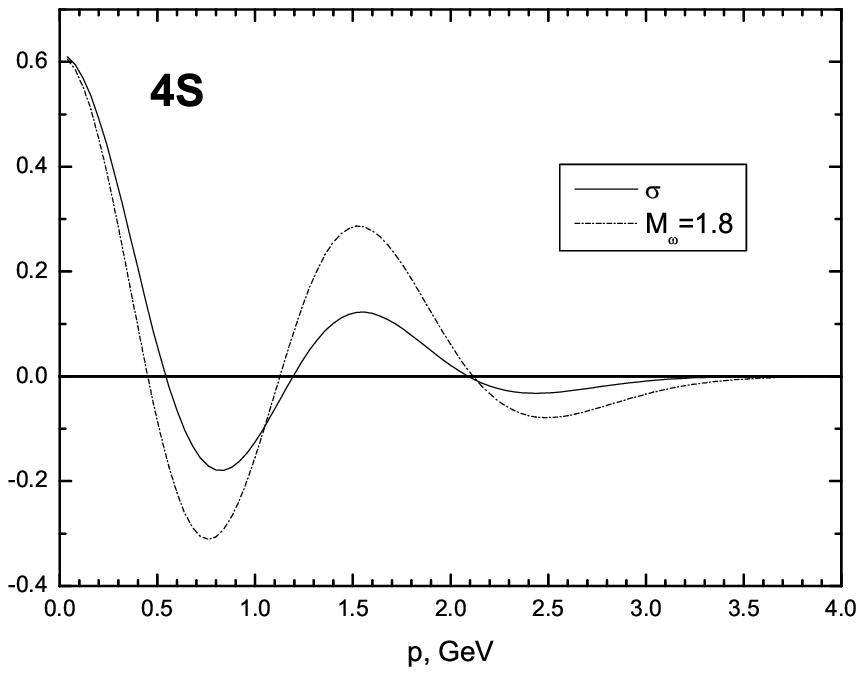}
\includegraphics[width= 6cm,height=5cm,keepaspectratio=true]{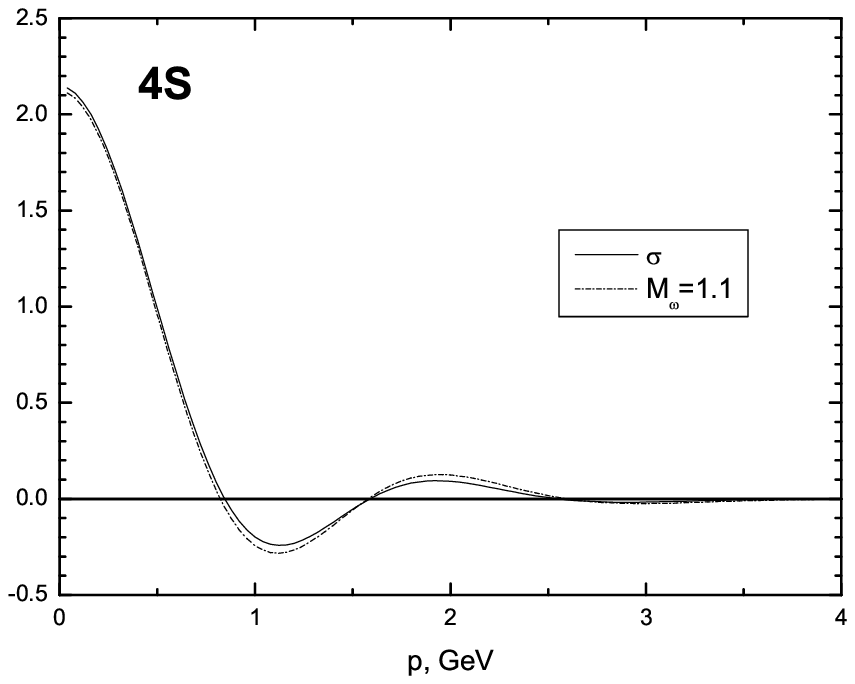}
\caption{The same as in Fig. 2, but for the  $4S (c\bar c)$ into $DD$ (left
panel) and $4S(b\bar b)$ into $BB$ (right panel).}}
\end{figure}

\begin{figure}[h]
\center{
\includegraphics[width= 6cm,height=5cm,keepaspectratio=true]{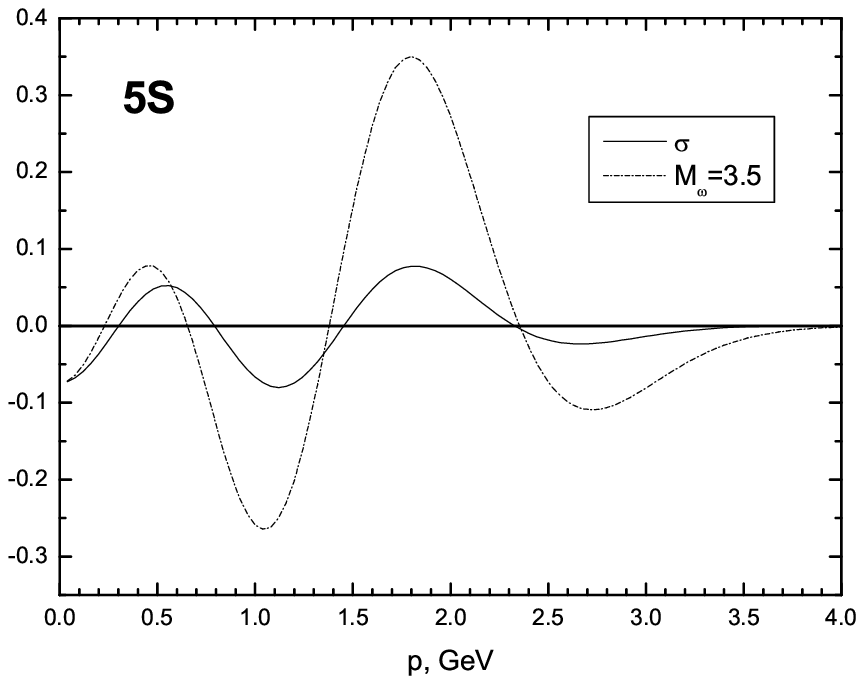}
\includegraphics[width= 6cm,height=5cm,keepaspectratio=true]{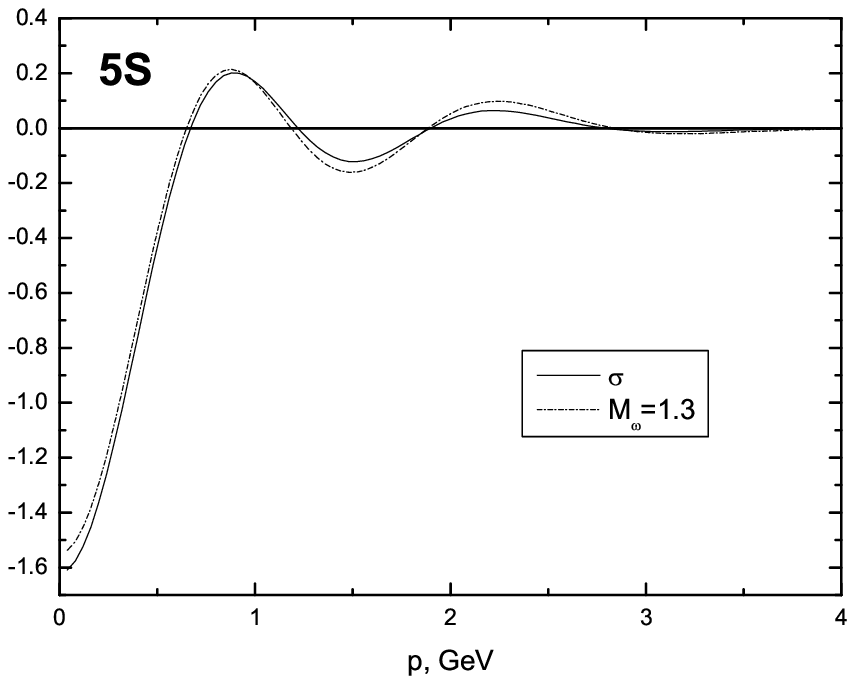}
\caption{The same as in Fig. 2, but for the  $5S (c\bar c)$ into $DD$ (left
panel) and $5S(b\bar b)$ into $BB$ (right panel).}}
\end{figure}

\section{Analytic and phenomenological study of unquenched spectra}

Our RSH was derived in Ref.~\cite{5} starting from the Wilson loop for the
$q\bar q$ system and using Nambu-Goto action for the corresponding string. In
the derivation  presence of additional quark loops was neglected (quenched
approximation), basing on the $1/N_c$ argument and additional
(phenomenological) numerical suppression of the quark-loop effects. It is the
purpose of the present Section to study these effects analytically and
phenomenologically, and compare them with lattice results in the forthcoming
Section.

The generating functional of heavy charges $Q\bar Q$, after
integrating over other quark-loops, has the form \be Z=\int DA\exp
\ \mathcal{L}_A W_{Q\bar Q} (A) \det (m_q+ \hat
D(A)),\label{12}\ee where $\mathcal{L}_A$ is the standard gluonic
action and $W_{Q\bar Q} (A)$ is the external (fixed) Wilson loop
of heavy quarks. The $\det$ term can be written in the path
integral form \cite{24}: \be \det (m_q+ \hat D(A)) =\exp\left [tr
\ln \left( \frac12 \int^\infty_0 \frac{ds}{s} (D^4z) e^{-K}
W_{q\bar q} (A) \right)\right],\label{13}\ee where $(D^4z)$ is the
path integration, $s$ is the proper time variable, $K=\frac14
\int^s_0 \left(\frac{dz_\mu(\tau)}{d\tau}\right)^2 d\tau$, and
$W_{q\bar q } (A)$ is the Wilson loop of sea quarks, while $tr$
implies summation over flavor indices and space-time coordinates.

Expanding in the number of sea-quark loops, one has the first correction term
\cite{24}: \be Z_{\rm 1 loop} = \int DA \exp \mathcal{L}_A \left( \frac12 tr
\left\{ \int^\infty_0 \frac{ds}{s} D^4 z) e^{-K} W_{q\bar q} (A) W_{Q\bar Q}
(A) \right\} \right).\label{14}\ee Integrating in (\ref{14}) over $DA$, one
obtains the effective one-loop partition function, \be Z_{\rm 1 loop} =
-\frac12 \int^\infty_0 \frac{ds}{s} (D^4z) e^{-K} \chi (W_{q\bar q}, W_{Q\bar
Q}),\label{15}\ee where $\chi$  is connected average of two loops, \be \chi
\equiv\lan W_{q\bar q} (A) W_{Q\bar Q} (A) \ran - \lan W_{q\bar q} (A) \ran
\lan W_{Q\bar Q} (A) \ran. \ee Properties of $\chi$ were studied in
\cite{25,26}, where it was shown that one can find a simple expression for
$\chi$ for small distances between minimal area surfaces of both Wilson loops,
\be \chi \cong \frac{1}{N_c^2} \exp (-\sigma S_\Delta)\label{17}\ee and
$S_\Delta$ is the minimal area of the surface connecting contours $C_1$ and
$C_2$ of Wilson loops $W_{Q\bar Q}$ and $W_{q\bar q}$, respectively, as shown
in Fig.7 One can see in Fig.7 that the width of the bands in $S_\Delta$ along
time direction is of the order of $ R_2 , R_3$, where $R_2, R_3$ are the radii
of intermediate mesons $(Q\bar q)$ and $(\bar Q q)$.

\begin{figure}[h]\center{
\includegraphics[height=8cm]{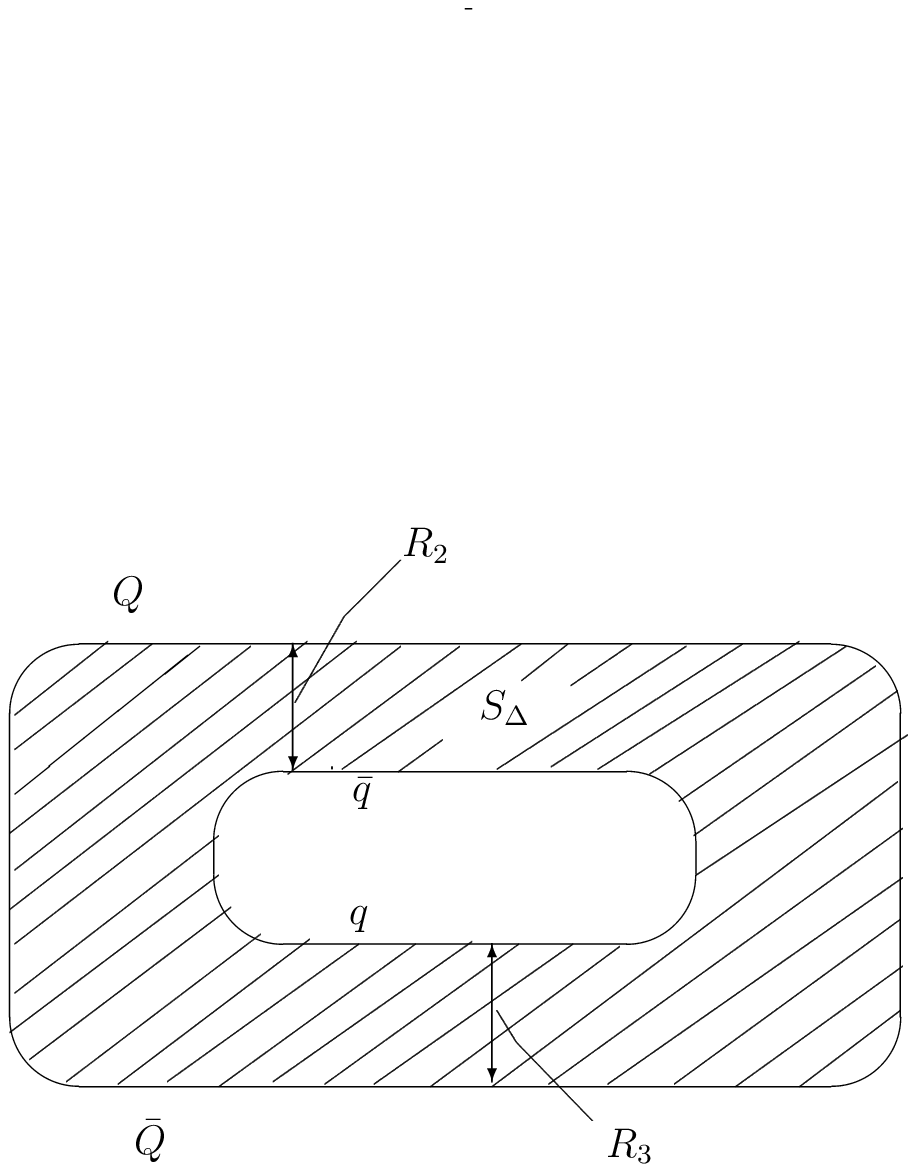}
 \caption{Connected  average of two Wilson loops, expressed via area
law for the surface $S_\Delta$ between two loops. Radii $R_2$ and
$R_3$ of two intermediate mesons $(Q\bar q)$ and $(\bar Q q)$ are
also shown.} }
\end{figure}

Several properties of the effective partition function (\ref{15}) can be
derived immediately:

1) The general property of the unquenching process: since the
integral $\int\frac{ds}{s}$ is obviously diverging at small $s$,
one needs a renormalization step, which means that the string
tension $\sigma$ in (\ref{17}) is the renormalized (by
unquenching) version of the quenched $\sigma$.

Expanding averaged static Wilson loop with sea quarks, one obtains \be \lan
W_{Q\bar Q|q\bar q}\ran  = \lan W_{Q\bar Q} \ran + \frac{1}{N_c} \llan W_{Q\bar
Q} \bar W_{q\bar q}\rran + \frac{1}{N^2_c}\llan W_{Q\bar Q} \bar W_{q\bar
q}\bar W_{q\bar q}\rran+...,\label{18}\ee where the bar over $W_{q\bar q}$
implies averaging over all paths, i.e. all contours of light quarks $C_{q\bar
q}$ with the weight defined by the Fock-Feynman-Schwinger  path integral, \be
\bar W_{q\bar q} = \int^\infty_{s_0} \frac{ds}{s} (D^4z) e^{-K} W_{q\bar q}
(C_{q\bar q}). \label{19}\ee Correspondingly, one can write each Wilson loop
and their products as (omitting correction terms independent of $T$). \be \lan
W_{Q\bar Q} \ran = \exp (-V_{Q\bar Q} (R) T)\label{20}\ee \be \lan W_{Q\bar Q}
\bar W_{q\bar q}\rran = \exp (-V_\Delta (R) T);\label{21}\ee \be \lan W_{Q\bar
Q} \bar W_{q\bar q} \bar W_{q\bar q} \rran = \exp (-V_{\Delta\Delta} (R)
T).\label{22}\ee Therefore the resulting $Q\bar Q$ interaction appears to be
dependent on $T$. Since $V_{Q\bar Q} (R) = \sigma_{\rm ren} R + V_{\rm GE}
(R)$, while $V_{\Delta\Delta}(R) < V_\Delta (R) < V_{Q\bar Q}(R)$ for large
enough $R$, with increasing $T$ the $Q\bar Q$ system will pass from the purely
confining regime $V_{Q\bar Q}$ to one-loop regime $V_\Delta$ and then to
two-loop regime $V_{\Delta\Delta}$ etc. In the next Section we shall show that
this type of transition was indeed observed on the lattice.

As to the form of $V_\Delta (R), V_{\Delta\Delta} (R)$ etc., one expects that
for $R< R_2 + R_3$, where $R_2, R_3$ are radii of lowest  $(Q\bar q), (\bar Q
q)$ states, the form of $V_\Delta (R)$ does not change, i.e. \be V_\Delta(R)
\approx \sigma_{\rm ren} R,~~ R<R_2+R_3\approx 1 {\rm fm}.\label{23}\ee

In case of the $c\bar c$ system $R_2=R_3= R_D\approx  R_{D^*} \approx 0.6$ fm.
The same is true for $b\bar b$ system with $R_B \approx 0.5$ fm.

In a similar way one can treat $V_{\Delta\Delta}$ and higher loop
terms. As a result one can predict that the static potential can
be defined from the sum (\ref{18}) in the $T$- independent way for
$R\la 1$ fm, \be V_{\rm static} (R) \approx V_{Q\bar Q} (R)
\approx  \sigma_{\rm ren} R, ~~R\la 1 {\rm fm}.\label{24}\ee

For $R>1.2$ fm the situation is complicated and static
$(T$-independent) potential cannot be defined in the strict sense,
as was discussed above. In this case another approach can be used,
namely, the expansion of the connected averages $\llan W_{Q\bar
Q}\bar W_{q\bar q}\rran$ in the series over intermediate
heavy-light meson states,  as was done  in  \cite{17}, and it is
equivalent to the  expansions in \cite{4}, \cite{16},
\cite{19,20,21,22}. In this way instead of $V_\Delta (R)$ one
defines the energy-dependent nonlocal interaction \be V_{121}
(\veq,\veq', E) = \sum_{n_2n_3} \int \frac{d^3\vep}{(2\pi)^3}
\frac{X_{n_2n_3} (\veq-\vep) X^+_{n_2n_3}
(\veq'-\vep)}{E-E_{n_2n_3} (\vep)}, \label{25}\ee where subscripts
1,2 refer to the channels $Q\bar Q$ and $(Q\bar q) (\bar Q q)$,
respectively, while $n_2, n_3$ are quantum numbers of the mesons
$(Q\bar q)$ and $(\bar Q q)$ with the wave functions $\psi_{n_2}$
and $\psi_{n_3}$, and \be X_{n_2n_3} (\ver) =\frac{M_\omega \eta
(\beta, \gamma)}{\sqrt{N_c}} \int \frac{d^3\veq}{(2\pi)^3}
e^{i\veq\ver} \psi_{n_2} (\veq) \psi_{n_3} (\veq).\label{26}\ee In
a similar way instead  of $V_{\Delta\Delta} (R)$, one defines the
interaction $V_{131}$ due to three-meson intermediate states.

As a result, the total Hamiltonian has the form \be H=H_{kin} + V_{Q\bar
Q}(\veR) + V_{121} (\veR, \veR', E) + V_{131} (\veR, \veR', E)
+....\label{27}\ee  As one can see, in (\ref{27}) the $T$-dependent interaction
of (\ref{20}), (\ref{21}), (\ref{22}) is replaced by the energy-dependent nonlocal interaction.

Let us underline general properties of the new Hamiltonian (for an
earlier discussion see \cite{31}).

i) For energies below all thresholds the interaction
$\sum_{n\geq2} V_{1n1} (\veR, \veR, E)$ is negative, which implies
attraction on average from all higher intermediate states. Hence
the linear potential in $V_{Q\bar Q} (R)$ is modified (flattened)
by inclusion of intermediate states. This attraction also persists
in some energy region above thresholds, where the real part of $
V_{1n1}$ is still negative.

ii) Due to strong reduction of overlap integrals of the type
$||\Psi_k (\veR) V_{1n1} $ $(\veR, \veR', E) \Psi_l (\veR')||$ (as
was discussed in previous Section, it is due to the string width
effect), the series $\sum_{n\geq 2} V_{1n1}$ is fast converging
and therefore only few terms are important.

Summarizing the effect of sea-quark loop on the $Q\bar Q$
interaction, one can say that there is no energy-independent (or
time-independent) universal local interaction which can describe
the dynamics of $Q\bar Q$ system in the unquenched case. If one
tries to simulate the effect of quark loops on the static $Q\bar
Q$ potential, then it should be an approximate local interaction,
which is close to linear potential $\sigma _{\rm ren} R$ for $R\la
1$ fm, and becomes softer (flattening) for larger $R$, which can
be approximated by making $\sigma_{\rm ren}$ the energy -- and $R$
-dependent.

Such kind of flattening potential was introduced in \cite{14} to describe high
excitations of light mesons and used latter in \cite{11,12,13} for higher
charmonium states. \be \tilde V_{Q\bar Q} (R) = \sigma (R) \cdot R;~~ \sigma
(R) = \sigma_0 \left[ 1- \gamma_0 \frac{\exp (\sqrt{\sigma_0} (r-R_1))}{B+\exp
(\sqrt{\sigma_0} (r-R_1))}\right].\label{28}\ee Here $R_1 \approx 1.2$ fm is
the distance, where the string can decay into two mesons, $\sigma_0 R_1 +
2M_Q\approx 2 M_{Q\bar q}, \sigma_0= 0.19$ GeV$^2$. Putting $\gamma_0=0.40$,
the best  description of radial excited light mesons was obtained in \cite{14}.
The modified potential $\tilde V_{Q\bar Q} (R)$, taken from \cite{14}, is shown
in Fig.8.

\begin{figure}
\epsfig{figure=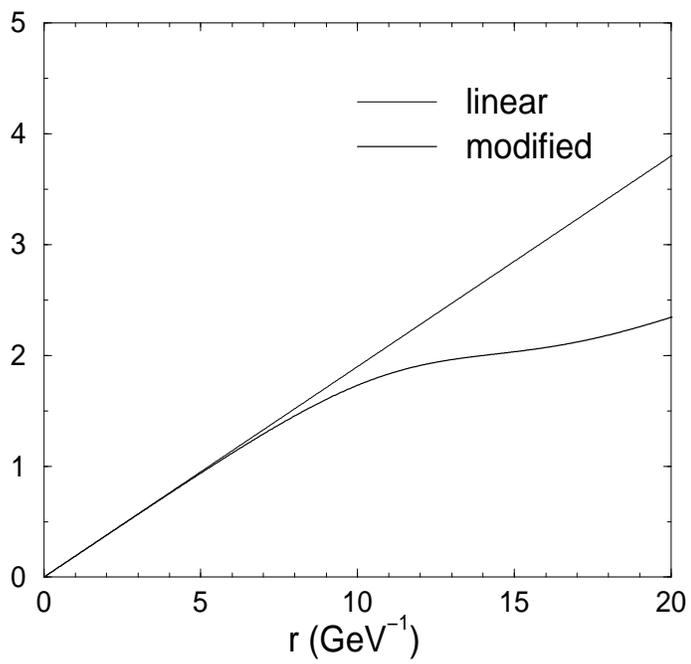,height=90mm,width=90mm} \caption{Modified potential with
parameters given in Eq.~(\ref{28}). For reference  simple linear potential with
$\sigma = 0.19$ GeV$^2$ is also plotted. \label{fig.1}}
\end{figure}

The resulting light meson masses, taken from \cite{14}, are
compared with experimental data in Table 1. For heavy quarkonia
the role of the flattening is less important. Below in Table 2 one
can see corresponding effect in charmonium levels, which is of the
order of several tens of MeV for high excitations.

Another types of flattening potentials were suggested in \cite{*}. One should,
however, be careful with the large $R$ behavior of their flattening potentials,
which     is bounded from above,  and therefore the  quarks $Q, \bar Q$ can
liberate themselves and this effect contradicts the physical picture in QCD,
when an unstable hadron decays into hadrons, but not into quarks.

\begin{table}
\caption{ Light meson masses $M_0(nL)$ (in GeV) from RSH with pure
linear potential and their masses $\tilde M_0(nL)$  in potential
with flattening. The corresponding flattening correction
$\delta_{\rm flat} $ (in MeV), the self-energy correction
$\Delta_{SE} =- \frac{12 \sigma}{\pi M_0}$, the string correction
$ \Delta_{str} = \frac{2 \sigma \lan r^{-1}\ran l (l+1)}{M^2_0}$,
and the resulting mass $M_{tot} = \tilde M_0 + \delta_{SE}
+\Delta_{str} + \delta_{GE}$ (where $\delta_{GE}$ is the
correction from the gluon-exchange potential) are given in GeV.}
\vspace{0.5cm}
\begin{tabular}{|r|r|r|r|r|r|c|}
\hline State& $\gamma=0$& $\gamma=0.4$& $\delta_{\rm flat}$
(MeV)&$\Delta_{SE}$& $\Delta_{str} $&  $M=\tilde M_0$\\
nL & $M_0(nL)$& $M_0(nL)$&flattening& &&\\\hline
1S& 1.347&1.335& -12& -0.510& 0& $0.725$\\
2S & 2.009& 1.944& -65&-0.342&0& $1.503$\\
1D&2.167&2.122&-45&-0317&-0.087&1.662\\
3S&2.512&2.300&-212&-0.274&0& 1.937\\
2D&2.615&2.428&-187&-0.263& -0.058 &2.052\\
4S& 2.931&2.569&-362&-0.235&0&2.252\\
3D&3.006&2.647&-359&-0.229&-0.043&2.322\\
1P&1.802&1.777&-25&-0.382&-0.074&0.071\\
2P&2.328&2.213&-115&-0.295&-0.030&0.068\\
1F&2.479&2.402&-77&-0.277&-0.113&0.048\\
3P&2.766&2.472&-294&-0.249&-0.020&0.065\\2F&2.876&2.606&-270&-0.239&-0.083&0.047\\
4P& 3.146&2.731&-415&-0.219&-0.015&0.062\\
3F&3.233&2.821&-412&-0.213&-0.064&0.046\\

\hline
\end{tabular}
\end{table}

\begin{table}
\caption{Comparison of  the single-channel and flattening potential results for
S,P,D states of charmonium with existing experimental data.}
 \vspace{0.5cm}
 \begin{center}

\begin{tabular}{|r|r|r|r|}
\hline
State  &   SC  &   flattening   &  exp    \\
\hline
1S&3.068&3.066&3.067\\
2S&3.678&3670&3.74(4)\\
3S&4.116&4.093&$  4.040(3^3S_1)$\\
4S& 4.482&4.424&$ 4.421(4^3S_1)$\\
5S &4.806& 4.670& ?\\
\hline

1P&3.488&3.484& 3.525 $(^1P_1)$\\
2P&3.954&3.940&$\sim 3.93 (^5P_2)$\\3P& 4.338&4.299& -\\
\hline

 1D &3.79&3.78&$3.77(1^3D_1)$\\
2D &4.189&4.165& 4.153(3)\\
3D& 4537&4.475&-\\

\hline
\end{tabular}
\end{center}
\end{table}

\section{Comparison to other approaches}

Here we compare our string decay picture with lattice data and other
approaches. On the lattice the topic of string breaking and $Q\bar Q$
interaction above inelastic threshold was actively explored during last decade
(for the first attempts see \cite{33} and \cite{34}). A way to determine the
static potential in unquenched case was  suggested in \cite{35} and several
spectra calculations, including sea-quark effects, were done in \cite{36}.
Recently careful studies of spectra of excited hadrons with open channels were
published in \cite{37} and \cite{38}, where the importance of inelastic
channels was stressed. The difficulty of existing lattice approaches is the
lack of the proper definition of a resonance state, which actually belongs  to
the continuous spectrum and requires either the continuous density description
or the use of the Weinberg Eigenvalue Method, described recently in the last
paper of Ref.~\cite{18}. The first approach is made possible by the use of the
finite volume, when continuous states are discretized and the resonance is
defined by the scattering phase \cite{33,34}. The second approach, to our
knowledge, was never used on the lattice. As to precise definition of the
resonance parameters on the lattice, from \cite{37,38} one can see that it
needs a lot of efforts and is expected in the nearest future.

It is worth saying that the  $Q\bar Q$ potential, calculated on
the lattice, is not sensitive to the effects of the  virtual sea
quarks at least for distances $R\la 1$ fm (for the latest
calculation see \cite{39}). This result is in agreement with our
discussion of the structure of the unquenched Wilson loop in the
previous Section.

\begin{table}
\caption{Comparison of calculated mass $M_{tot}$ (in GeV) with
experimental data.}
\begin{center}
\begin{tabular}{|r|r|r|r|}
\hline
$M_{tot}$ & Theory& Exp.&  $M_{cog}(th)$\\

\hline

$M(\rho(1S))$& 0.749&$\rho(0.775)$& 0.666\\

$M(\rho(2S))$& 1.519&$\rho(1.465)$& 1.479\\

$M(\rho(3S))$& 1.937&$\rho(1900)?$&1.849\\

$M(\rho(4S))$& 2.252&$\rho(2150)?$&2.166\\

$\bar M(1P)$& 1.25&$a_1(1230)$&\\
&&$f_1(1282)$&\\

$\bar M(2P)$& 1.82&$a_1(1647)? a_2(1732)?$&\\
&&$f_1(1815)?$&\\

$\bar M(3P)$& 2.14&$f_2(2157)$ mixing for $3^3P_2$&\\

$\bar M(4P)$&2.435&&1.65\\

$\bar M(1D)$& 1.66&$\rho_3(1690), \rho(1720)$&\\
&& mixing for $2^3S_1-1^3D_1$&\\

$\bar M(2D)$& 2.05&$\rho_3(1990)?$& 1.989\\
$\bar M(3D)$& 2.32&$\rho_3(2250)?, \rho_5(2330)?$&2.249\\
&&$(3D-2G$ mixing(?))&\\
$\bar M(1F)$& 1.96&$a_4(2000), f_4(2018)$&\\
$\bar M(2F)$& 2.24&$ f_4(2300)$&\\

$\bar M(3F)$& 2.50&&\\

\hline
\end{tabular}
\end{center}

\end{table}



\begin{thebibliography}{99}

\bibitem{1}
J.~M.~Richard, Phys. Lett. B {\bf 100}, 515 (1980); ibid {\bf 95},
299 (1980); D.~P.~Stanley and D.~Robson, Phys. Rev. Lett. {\bf
45}, 235 (1980); Phys. Rev. D {\bf 21}, 3180 (1980); P.~Cea,
G.~Nardulli, and G.~Preparata, Z.Phys. C {\bf 16}, 135 (1982);
J.~Carlson, J.~Kogut, and V.~R.~Pandharipande, Phys. Rev. D {\bf
27}, 233 (1983); J.~L.~Basdevant and S.Boukraa, Z. Phys. C {\bf
28}, 413 (1983).

\bibitem{2}
N.~Isgur and S.~Godfry, Phys. Rev. D {\bf 32}, 189 (1985).


\bibitem{3}
S.~Capstick and N.~Isgur, Phys. Rev. D {\bf 34}, 2809 (1986).

\bibitem{4}
E.~Eichten, K.~Gottfried, T.~Kinoshita, K.~D.~Lane, and T.~M.~Yan,
  Phys. Rev. D {\bf 17}, 3090 (1978) (Erratum:ibid. D {\bf 21}, 313
(1980)), ibid. D {\bf 21}, 203 (1980); Phys. Rev. Lett. {\bf 36},
500 (1976).



\bibitem{5}
A.~Yu.~Dubin, A.~B.~Kaidalov, and Yu.~A.~Simonov, Phys. Lett. B
{\bf 323}, 41 (1994); Phys. At. Nucl.  {\bf 56}, 1745 (1993);
Nuovo Cim. A {\bf 107}, 2499 (1994); E.~L.~Gubankova and
A.~Yu.~Dubin, Phys. Lett. B {\bf 334}, 180 (1994).

\bibitem{5'}
Yu.~A.~Simonov, in ``QCD: Perturbative or Nonperturbative ?" Ed.
by L.~Ferreira, P.~Nogueira, and J.~I.~Silva-Marcos ( World Sci.,
Singapore (2001), p.60; arXiv: hep-ph/9911237.


\bibitem{6}
Yu.~S.~Kalashnikova, A.~V.~Nefediev, and Yu.~A.~Simonov, Phys.Rev.
D {\bf 64}, 014037  (2001).


\bibitem{7}
Yu.~A.~Simonov, Phys. Lett. B {\bf 515}, 137  (2001);  A.~Di~
Giacomo and Yu.~A.~Simonov, Phys. Lett. B {\bf 595}, 368 (2004).



\bibitem{8}
A.~M.~Badalian and B.~L.~G.~Bakker,  Phys. Rev. D {\bf 66}, 034025
(2002).

\bibitem{9}
V.~L.~Morgunov, A.~V.~Nefediev, and Yu.~A.~Simonov, Phys. Lett. B
{\bf 459}, 653 (1999).


\bibitem{10}
Yu.~A.~Simonov and V.~I.~Shevchenko,  Adv. High Energy,  {\bf
2009}, 873051 (2009); arXiv: 0902.1405 [hep-ph]; Yu.~A.~Simonov,
arXiv: 1003.3608 [hep-ph].

\bibitem{11}
A.~M.~Badalian, Phys. At. Nucl. {\bf 74}, 1375 (2011); arXiv:
1011.5580 [hep-ph].

\bibitem{12}
A.~M.~Badalian, B.~L.~G.~Bakker, and I.~V.~Danilkin, Phys. Rev. D
{\bf 81}, 071502 (2010).

\bibitem{13}
A.~M.~Badalian and  B.~L.~G.~Bakker, Phys. Rev. D {\bf 84}, 034006
(2011); A.~M.~Badalian, B.~L.~G.~Bakker, and I.~V.~Danilkin, Phys.
At. Nucl. {\bf 74}, 631 (2011).

\bibitem{14}
A.~M.~Badalian, Phys. At. Nucl. {\bf 66}, 1342 (2003);
A.~M.~Badalian, B.~L.~G.~Bakker, and Yu.~A.~Simonov, Phys. Rev. D
{\bf 66}, 034006 (2002).

\bibitem{15}
S.~M.~Fedorov and Yu.~A.~Simonov, JETP Lett. {\bf 78}, 57 (2003);
hep-ph/0306216 [hep-ph].

\bibitem{16}
E.~van~Beveren, C.~Dullemond, and G.~Rupp, Phys. Rev. D {\bf  21},
772 (1980), E.~van~Beveren, Z.Phys. C {\bf 17}, 135 (1983);
E.~van~Beveren, G.~Rupp, T.~A.~Rijken, and C.~Dullemond, Phys. Rev
D {\bf 27}, 1527 (1983); S.~Jacobs, K.~J.~Miller, and
M.~G.~Olsson, Phys. Rev. Lett. {\bf 50}, 1181 (1983);  G.~Fogli
and G.~Preparata, Nuovo Cimento, A {\bf 48}, 235 (1978);
A.~C.~Maciel and J.~Paton, Nucl. Phys. B {\bf 181}, 277 (1981);
N.~A.~ T\"{o}rnqvist, Ann. Phys. (N.Y.) {\bf 135}, 1 (1978);
M.~Ross and N.~A.~T\"{o}rnqvist, Z. Phys. {\bf C 5}, 205 (1980),
N.~A.~T\"{o}rnqvist,  Phys.  Rev. Lett. {\bf 49}, 624 (1982);
Nucl. Phys. B {\bf 203}, 268 (1982).


\bibitem{17}
Yu.~A.~Simonov, Phys. At. Nucl. {\bf 71}, 1048 (2008),
arXiv;0711.3626; Yu.~A.~Simonov, A.~I.~Veselov, Phys. Rev. D {\bf
79}, 034024 (2009); I.~V.~Danilkin and  Yu.~A.`Simonov, Phys. Rev.
D {\bf 81}, 074027 (2010).

\bibitem{18}
L.~Micu, Nucl. Phys. B {\bf 10}, 521 (1969); A.~Le~Yaouanc,
L.~Olivier, O.~Pene, and J.~Raynal, Phys. Rev. D {\bf 8}, 2223
(1973); ibid. D {\bf 9}, 1415 (1974); D {\bf 11}, 1272 (1975); D
{\bf 21}, 182 (1980).




\bibitem{19}
N.~Isgur and  J.~Paton, Phys. Rev.  D {\bf 31}, 2910 (1985);
H.~Blundell and S.~Godfrey, Phys. Rev. D {\bf 53}, 3700 (1996);
T.~Barnes, F.~E.~Close, P.~R.~Page, and E.~S.~Swanson, Phys. Rev.
D {\bf 55}, 4157 (1997);  P.~K.~Page, Nucl. Phys. B {\bf 446}, 189
(1995); P.~Geiger and E.~S.~Swanson, Phys. Rev.  D {\bf 50}, 6855
(1994); H.~Q.~Zhou, R.~G.~Ping, and B.~S.~Zou, Phys. Lett. B {\bf
611}, 123 (2005); X.~H.~Guo, H.~W.~Ke, X.~Q.~Li, X.~Liu, and
S.~M.~Zhao, Commun. Theor. Phys. {\bf 48}, 509 (2007); J.~Lu,
W.~Z.~Deng, X.~L.~Chen, and S.~L.~Zhu,  Phys. Rev. D  {\bf 73},
054012 (2006).


\bibitem{20}
E.~S.~Ackleh, T.~Barnes, and E.~S.~Swanson, Phys. Rev. D {\bf 54},
6811 (1996).


\bibitem{21}
K.~Heikkil\'{a}, N.~A.~T\"{o}rnqvist, and S.~Oho, Phys. Rev. D
{\bf 29}, 110 (1984); Yu.~S.~Kalashnikova,  Phys. Rev. D {\bf 72},
034010 (2005); M.~K.~Pennington and D.~J.~Wilson, Phys. Rev. D
{\bf 76}, 077502 (2007); T.~Barnes and E.~S.~Swanson, Phys. Rev. C
{\bf 77}, 055206 (2008);  S.~Godfrey and R.~Kokoski,  Phys. Rev. D
{\bf 43}, 1679 (1991);   F.~E.~Close and E.~S.~Swanson, Phys. Rev.
 D {\bf 72}, 094004 (2005);  E.~S.~Swanson,   J. Phys.  G {\bf
31}, 845 (2005);  D.~S.~Hwang and D.~W.~Kim, Phys. Lett. B {\bf
601}, 137 (2004);  E.~J.~Eichten, K.~Lane, and C.~Quigg, Phys.
Rev. D {\bf 69}, 094019 (2004);  C.~Hanhart, Yu.~S.~Kalashnikova,
A.~E.~Kudryavtsev, and A.~V.~Nefediev, Phys. Rev. D {\bf 76},
034007 (2007); Yu.~S.~Kalashnikova, AIP Conf. Proc. {\bf 892}, 318
(2007); Yu.~S.~Kalashnikova, Phys. Rev. D {\bf 72}, 034010 (2005);
C.~Amsler and N.~A.~T\"{o}rnqvist, Phys. Rept. {\bf 389}, 61
(2004).



\bibitem{22}
P.~Geiger and N.~Isgur, Phys. Rev. D {\bf 41}, 1595 (1990), ibid D
{\bf 44}, 799 (1991), ibid D {\bf 47}, 5050 (1993).

\bibitem{24}
Yu.~A.~Simonov, Phys. Rev. D {\bf 84}, 065013 (2011).

\bibitem{25}
Yu.~A.~Simonov, Phys. Usp. {\bf 39}, 313 (1996), [hep-ph/9709344];
D.~S.~Kuzmenko, V.~I.~Shevchenko, and Yu.~A.~Simonov, Phys. Usp.
{\bf 104}, 3 (2004).

\bibitem{26}
L.~Del~Debbio, A.~Di~Giacomo,  and  Yu.~A.~Simonov, Phys. Lett. B
{\bf 332}, 111 (1994);  N.~Cardoso, M.~Cardoso, and P.~Bicudo,
 Phys. Lett. B {\bf 710}, 343 (2012).

\bibitem{27}
R.~W.~Haymaker, V.~Singh, Y.~Peng, and J.~Woisiek, Phys. Rev. D
{\bf 53}, 389 (1996);  P.~Bicudo, N.~Cardoso, and  M.~Cardoso,
arXiv: 1010.3870.

\bibitem{28}
H.~G.~Dosch, Phys. Lett. B {\bf 190},  177 (1987);   H.~G.~Dosch
and Yu.~A.~Simonov, Phys. Lett. B {\bf 205},  339 (1988);  A.~Di~
Giacomo, H.~G.~Dosch, V.~I.~Shevchenko, and  Yu.A.Simonov, Phys.
Rept. {\bf 372}, 319 (2002).

\bibitem{29}
G.~S.~Bali, N.~Brambilla, and  A.~Vairo, Phys. Lett. B {\bf 421},
265 (1998); Y.~Koma, M.~Koma, Nucl. Phys. B {\bf 769}, 79 (2007).

\bibitem{28*}
Yu.~A.~Simonov, Phys. At. Nucl. {\bf 71}, 1048 (2008);
Yu.~A.~Simonov and A.~I.~Veselov, Phys. Rev. D {\bf 79}, 034024
(2009).

\bibitem{29*}
Yu.~A.~Simonov and A.~I.~Veselov, Phys. Lett. B{\bf 671}, 55
(2009).

\bibitem{30*}
I.~V.~Danilkin, V.~D.~Orlovsky, and Yu.~A.~Simonov, Phys. Rev. D
{\bf 85}, 034012 (2012).

\bibitem{31*}
T.~Barnes, S.~Godfrey, and E.~S.~Swanson, Phys. Rev. D {\bf 72},
054026 (2005).

\bibitem{31}
A.~M.~Badalian, L.~P.~Kok, M.~I.~Polikarpov, and Yu.~A.~Simonov,
Phys. Rept. {\bf 82}, 31 (1982).

\bibitem{*}
Y.~B.~Ding, K.~T.~Chao, and D.~H.~Qin, Chin. Phys. Lett. {\bf 10},
460 (1993); Phys. Rev. D {\bf 51}, 5064 (1995); B.~Q.~Li, C.~Meng,
and K.~T.~Chao, Phys. Rev. D  {\bf 80}, 014012 (2009).

\bibitem{33}
M.~L\"{u}scher, Comm. Math.Phys. {\bf 104}, 177 (1986); ibid {\bf
105}, 153 (1986);  Nucl. Phys. B {\bf 354}, 531 (1991); ibid. B
{\bf 364}, 237 (1991).

\bibitem{34}
C.~Mc~Neil and C.~Michael (UKQCD), Phys. Lett. B {\bf 556}, 177
(2003); C.~Michael, Eur. Phys. J. A {\bf 31}, 793 (2007);
Z.~Prkacin, G.~S.~Bali, Th.D\"{u}ssel et al, PoS
\textbf{LATTICE2005}, 308 (2005) [hep-lat/0510051].

\bibitem{35}
M.~Donnellan, F.~Knechtli, B.~Leder, and R.~Sommer, Nucl. Phys. D
{\bf 849}, 45 (2011).

\bibitem{36}
T.~Burch \textit{et al.}  [Fermilab Lattice and MILC Collab.],
Phys. Rev. D {\bf 81}, 034508 (2010);  E.~Follana \textit{et al.}
[HPQCD and UKQCD Collab.], Phys. Rev. D {\bf 75}, 054502 (2007);
Y.~Namekawa \textit{et al.}  [PACS-CS Collab.], PoS
\textbf{LATTICE2008}, 121 (2008);  Y.~Namekawa \textit{et al.}
[PACS-CS Collaboration], Phys. Rev. D {\bf 84}, 074505 (2011);
G.~S.~Bali \textit{et al.},   PoS \textbf{LATTICE2011}, 135
(2011);   S.~M.~Ryan [Hadron Spectrum Collab.], PoS
\textbf{LATTICE2010}, 124 (2010).


\bibitem{37}
G.~S.~Bali, S.~Collins, and  C.~Ehmann, Phys. Rev. D {\bf 84},
094506 (2011).

\bibitem{38}
C.~Morningstar \textit{et al.}, AIP Conf. Proc. {\bf 1441}, 290
(2012); arXiv:09.0308 [hep-lat].


\bibitem{39}
T.~Kawanai and S.~Sasaki [Tokyo U., RIKEN BNL, and Brookhaven
Collab.], Phys. Rev. D {\bf 85}, 091503 (2012).

\end{thebibliography}
\end{document}